\newcommand{\kms}{\rm ~km~s^{-1}}
\newcommand{\ergs}{\rm ~erg~s^{-1}}

\documentclass[preprint]{aastex}

\begin{document}

\title{IMPLICATIONS OF THE X-RAY PROPERTIES OF PULSAR NEBULAE}
\author{Roger A. Chevalier}
\affil{Department of Astronomy, University of Virginia, P.O. Box 3818}
\affil{Charlottesville, VA 22903; rac5x@virginia.edu}


\begin{abstract}

A plausible model for the Crab Nebula is one in which a particle dominated,
highly relativistic wind from the pulsar passes through a shock front
in which the particles attain a power law energy distribution.
The electrons and positrons lose energy radiating synchrotron emission.
Here, a one zone version of the model is developed and applied to
observations of  X-ray pulsar nebulae.
Efficient conversion of pulsar spin-down power to
X-ray luminosity   is expected 
if the observed electrons
are in the synchrotron cooling regime and their energy  spectrum
is similar to that in the Crab Nebula, which has 
$N(\gamma)\propto \gamma^{-2.2}$, where $\gamma$ is the particle Lorentz factor.
In this case, the relation between 
X-ray luminosity and pulsar spin-down power depends only weakly on
the model parameters.
The dependence is stronger for a steeper particle spectrum, as appears to
be present in N157B, and the efficiency of X-ray production can be lower.
If the electrons are not in the cooling regime, the X-ray luminosity can
be  low, as appears to be the case for the compact nebula
around the Vela pulsar.
Slow cooling is likely to be caused by a sub-equipartition magnetic field in
the radiating region.
Observations can place constraints on the uncertain physics of relativistic
MHD shocks.  
The model is related to those developed for gamma-ray burst afterglows.

\end{abstract}


\section{INTRODUCTION}

Beginning with the Crab Nebula, extended sources of X-ray synchrotron
emission have been observed around a number of pulsars.
The power for the relativistic particles and magnetic fields is thought to
be the spin-down power of the pulsar, $\dot E$.
In cases where  $\dot E$ can be
measured, there is a relation between the X-ray luminosity, $L_x$,
and $\dot E$.
On the basis of $\it Einstein$ data, Seward  \& Wang (1998)
found a relation of the form 
$\log{L_x}=1.39\log{\dot E}-16.6$,
where $L_x$ is the X-ray luminosity in the 0.2--4 keV range and
$\dot E$ is the pulsar spin-down power in ergs s$^{-1}$.
The X-ray luminosity of the central pulsar is included in addition to
 the extended nebula.
On the basis of $\it ASCA$ data, Kawai, Tamura, \& Shibata (1998)
found the relation 
$\log{L_x}=(33.42\pm 0.20)+(1.27\pm 0.17)\log{(\dot E/10^{36})}$,
where $L_x$ is now the nebular luminosity in the 1--10 keV range.
Expressions like these have been widely used to estimate $\dot E$ in
X-ray synchrotron nebulae where a pulsar has not been detected.

The objects considered by Kawai et al. (1998) cover a range of $\dot E$
of $4\times 10^{33}-5\times 10^{38}\ergs$ and include symmetric objects
like the Crab Nebula, which may be interacting with supernova ejecta,
and the CTB 80, which appears to be a bow shock nebula interacting
with  a supernova remnant.
A compelling model for the optical/X-ray properties of the Crab
Nebula was developed by Rees \& Gunn (1974), Kundt \& Krotscheck (1980), and
Kennel \& Coroniti (1984a, 1984b; hereafter KC84a and KC84b).
In this model, the central pulsar generates a highly relativistic,
particle dominated wind that passes through a shock front and
decelerates to match the expansion velocity set by the outer nebula.
The electrons and positrons in the wind acquire a power law energy
spectrum in the shock front and radiate synchrotron emission in the
downstream region.
KC84b showed that this model is able to
account for the basic high energy properties of the Crab, including
the position of the wisps, the spectrum of the nebula, and the
size of the nebula at different wavelengths.
They were able to assume a steady state flow because the particle lifetimes
to synchrotron losses are less than the age of the Crab Nebula.
The model parameters are the Lorentz factor of the wind, $\gamma_w$,
the energy spectral index of the particles accelerated at the shock
front, $p$, and the magnetization parameter of the wind, $\sigma$, which
determines the magnetic field strength in the extended nebula.

Here, I develop a simplified version of the shocked wind model for 
the Crab Nebula which captures the essential features for estimating
the X-ray luminosity and spectrum (\S~2).
The model can be compared to the properties of other pulsar nebulae
in order to investigate the implications for the model parameters (\S~3).

\section{X-RAY EMISSION MODELS}

The model developed here is similar to that of
 KC84a and KC84b, but is a one zone model that
does not attempt to account for the spatial structure of the nebula.
The pulsar spin-down energy is assumed to go into a highly relativistic
wind with Lorentz factor $\gamma_w$.
As noted by KC84a, the magnetization parameter
in the wind, $\sigma$, must be small ($<0.1$) in order to have a high
efficiency of wind power into synchrotron radiation.
With these approximations, we have
\begin{equation}
\dot E\approx 4\pi n_1 \gamma_w^2 r_s^2 m c^3,
\end{equation}
where $n_1$ is the proper density in the wind just before the shock front,
$r_s$ is the shock wave radius, $m$ is the electron mass, and $c$ is
the speed of light.

At the shock front, the wind particles acquire a power law energy distribution
of the form $N(\gamma)\propto \gamma^{-p}$ for $\gamma \ge \gamma_m$,
where $n=\int N(\gamma)d\gamma$ is the density of relativistic particles
and $\gamma$ is the particle Lorentz factor.
I assume that $p>2$ and that the particle spectrum extends to higher
energies than those of the X-ray emitting particles.
In the first model,
the particles responsible for X-ray emission rapidly lose their energy to 
synchrotron radiation so there is a steady state.
The balance between injection from the shock front and synchrotron losses
leads to a constant number of emitting particles.
The basic simplification made here is that the emitting region can be
treated as one zone.
The energy density in the zone is approximately determined by the shock
jump conditions.
The particle pressure behind a highly relativistic shock wave is
$2n_1mc^2\gamma_w^2/3$ and the dynamic pressure is $n_1mc^2\gamma_w^2/3$
(KC84a).
The flow  decelerates downstream from the shock front, so
that $n_1mc^2\gamma_w^2$ is an estimate of the pressure and
$3n_1mc^2\gamma_w^2$ is an estimate of the energy density.
I assume that the energy density in the emitting region
is divided between a fraction
$\epsilon_e$ in particles and a fraction $\epsilon_B$ in the magnetic field.
These parameters, which sum to unity, replace the wind magnetization
parameter, $\sigma$, of KC84a.
In the flow model of KC84a, $\sigma$ is small but the compressive flow 
downstream from the shock front builds up the magnetic field to close
to equipartition.
The behavior of the magnetic field in the downstream region is not
well understood.
The toroidal field structure of the KC84a model is likely to be
unstable (Begelman 1998), and polarization observations of the synchrotron
emission in the Crab Nebula show that the field structure is complex.
In the relativistic blast wave model for gamma-ray burst afterglows,
some form of magnetic field enhancement in the postshock region is
required and it is often characterized by an efficiency factor
$\epsilon_B$ like that defined above (e.g., Sari, Piran, \& Narayan 1998).

With these assumptions, the magnetic field in the emitting region is
\begin{equation}
B=\left(6\epsilon_B \dot E\over r_s^2 c\right)^{1/2}.
\end{equation}
The postshock density in the shock frame is $3\gamma_w n_1$ for a
strong relativistic shock, so that the postshock particle distribution
function is $N(\gamma)=(p-1)3\gamma_w\gamma_m^{p-1}n_1\gamma^{-p}$.
The expression for the energy density then leads to
\begin{equation}
\gamma_m =\left(p-2 \over p-1\right) \epsilon_e\gamma_w.
\end{equation}

The number of radiating particles at a particular $\gamma$, 
$\cal N(\gamma)$, is determined by a balance
between the  rate at which particles are injected at the shock
front, $\dot{\cal N}(\gamma)$, and synchrotron losses.
The radiation power of an electron with Lorentz factor $\gamma\gg 1$ is
\begin{equation}
P(\gamma)={4\over 3}\sigma_T c \gamma^2 {B^2\over 8\pi}
\equiv \beta B^2\gamma^2,
\end{equation}
where $\sigma_T$ is the Thompson cross section and $\beta=1.06\times 10^{-15}$
cm$^3$ s$^{-1}$.
The total rate at which particles  are injected
is $\dot E/(\gamma_w mc^2)$, so
that
\begin{equation}
\dot{\cal N}(\gamma)=(p-1)\gamma_m^{p-1}(\gamma_w mc^2)^{-1}
\dot E \gamma^{-p}
\end{equation}
for the spectrum determined above.
The balance between injection and synchrotron losses can be expressed as
\begin{equation}
{1\over mc^2} {\partial {\cal N}\beta B^2\gamma^2\over \partial\gamma}
={\dot{\cal N}(\gamma)}
\label{evon}
\end{equation}
with the solution
\begin{equation}
{\cal N}(\gamma)=\gamma_m^{p-1}(\gamma_w\beta B^2)^{-1}\dot E\gamma^{-(p+1)}.
\end{equation}

I assume that the electron energy is radiated at its critical frequency
\begin{equation}
\nu(\gamma)=\gamma^2 {q_eB\over 2\pi mc}\equiv \xi \gamma^2 B,
\end{equation}
where $q_e$ is the electron charge and $\xi =2.80\times 10^6$ in cgs units.
The luminosity of radiating particles in the range $\gamma$ to
$\gamma+d\gamma$ is $\beta B^2\gamma^2 {\cal N}(\gamma)$, which leads to
a luminosity per unit frequency
\begin{equation}
L_{\nu}={1\over 2}\left(p-2\over p-1\right)^{p-1}
\left(6\xi^2\over c\right)^{(p-2)/4} \epsilon_e^{p-1}\epsilon_B^{(p-2)/4}
\gamma_w^{p-2} r_s^{-(p-2)/2}\dot E^{(p+2)/4}\nu^{-p/2}.
\label{lum}
\end{equation}
This expression is valid provided the X-ray 
frequency of observation, $\nu_{X}$, is greater 
than the cooling frequency $\nu_c$
and $\nu_m$, corresponding
to $\gamma_m$, and is less than $\nu_{max}$,
corresponding to the maximum $\gamma$ in the power law spectrum.
The cooling frequency corresponds to the $\gamma$ at which the electrons
are able to radiate their energy in the age of the wind nebula, $t$, and is
\begin{equation}
\nu_c={\xi\over B^3}\left(mc^2\over\beta t\right)^2
=\xi r_s^3\left(mc^2\over \beta t\right)^2 
\left(c\over 6\epsilon_B \dot E\right)^{3/2}.
\end{equation}
In the case of a bow shock nebula, the flow time past the nebula
may be the relevant value of $t$ rather than the total remnant age.
The other critical frequency is
\begin{equation}
\nu_m=\left(p-2 \over p-1\right)^2 {\epsilon_e^2\gamma_w^2\xi\over r_s}
\left(6\epsilon_B \dot E\over c\right)^{1/2}.
\end{equation}

The steady state assumption requires that the observing frequency, $\nu_o$,
be $>\nu_c$.
If $\nu_c<\nu_o < \nu_m$, then $\dot{\cal N}(\gamma)=0$ in eq. (\ref{evon}),
which has the solution ${\cal N}(\gamma)\propto \gamma^{-2}$.
This yields $L_{\nu}\propto \nu^{-1/2}$ (see also Sari et al. 1998).
The constant of proportionality is determined by joining the solution
onto the $\nu_o > \nu_m$ solution.

Observations of the compact nebula around the Vela pulsar
indicate $\nu_c>\nu_X$ and a different model is needed (see \S~3).
The Vela case is apparently a bow shock nebula created by the
proper motion of the pulsar (Markwardt \& \"Ogelman 1998).
The nebula age is then $t\approx r_s/v_p$, where $v_p$ is the
transverse pulsar velocity.
Because $t$ is less than the time since the birth of the pulsar,
the assumption of constant $\dot E$ is adequate.
The number of radiating particles is
${\cal N}(\gamma)=
\dot{\cal N}(\gamma)t=(p-1)\gamma_m^{p-1}(\gamma_w mc^2)^{-1}
\dot E r_s v_p^{-1}\gamma^{-p}$ and the luminosity is
\begin{equation}
L_{\nu}={1\over 2}{(p-2)^{p-1}\over (p-1)^{p-2}}
{6^{(p+1)/4}\over mc^{(p+9)/4}} \beta \xi^{(p-3)/2}\epsilon_e^{p-1}
\epsilon_B^{(p+1)/4}r_s^{-(p-1)/2}v_p^{-1}\gamma_w^{p-2}
\dot E^{(p+5)/4}\nu^{-(p-1)/2},
\label{lum2}
\end{equation}
which yields a lower luminosity than eq. (\ref{lum}) because of
the limited number of radiating electrons.

\section{THE CRAB AND OTHER PULSAR NEBULAE}

Detailed models have been produced for the Crab Nebula, so it provides
a point of reference for the approximate model developed here.
The X-ray spectrum of the Crab is a power law with photon index
2.1 (e.g., Pravdo \& Serlemitsos 1981), which corresponds to $p=2.2$ in the
present cooling model.
With this value of $p$, equation (\ref{lum}) becomes
\begin{equation}
L_{\nu}=0.084\epsilon_e^{1.2}\epsilon_B^{0.05}\gamma_w^{0.2}r_s^{-0.1}
\dot E^{1.05}\nu^{-1.1},
\label{lcrab}
\end{equation}
where cgs units are used.
The values $r_s=3\times 10^{17}$ cm and $\dot E=5\times 10^{38}\ergs$
are determined by observations of the Crab Nebula and its pulsar (see KC84b).
The efficiency factors, constrained by $\epsilon_e+\epsilon_B=1$, are not
well determined.
In the model of KC84a, the particles and magnetic field come into approximate
equipartition ($\epsilon_e=\epsilon_B=0.5$), although the field is initially
weaker.
The need for a particle dominated shock wave suggests that $\epsilon_e\ge 0.5$.
The quantity $\epsilon_e^{1.2}\epsilon_B^{0.05}$ is 0.42 in equipartition,
rises to 0.81 for $\epsilon_B=0.05$, and slowly drops to 0.63 for
$\epsilon_B=10^{-4}$; there is little sensitivity to this parameter.
I set it to 0.5.
The value of $\gamma_w$ depends on features in the nebular spectrum and
cannot be determined just from X-ray observations.
KC84b use $\gamma_w=3\times 10^6$ in their Fig. 13.
When these parameter values are substituted into eq. (\ref{lcrab}), the
result at $\nu=10^{18}$ Hz is $\nu L_{\nu}=1.0\times 10^{37}\ergs$,
which compares well to the detailed model  presented by KC84b
in their Fig. 13.
The model $0.2-4$ keV luminosity is $2.6\times 10^{37}\ergs$.

Although the luminosities agree, there are differences between the models.
In the present one zone model, the synchrotron burn-off leads to a steepening
of the frequency spectral index by 0.5, the standard result.
In the model of KC84b, the particles initially move out through a region
in which $B\propto r$ if the wind magnetization parameter is small.
If the synchrotron burn-off occurs in this region, the index steepens
by $(p+7)/18$, which is $>0.5$ for $p>2$.
The additional steepening occurs because at a high frequency, the particles
typically radiate in a lower magnetic field and higher $\gamma$ particles
are observed.
The fact that there are fewer of these particles steepens the spectrum.
However, for $p=2.2$, the steepening is by 0.51, very close to
the one zone model.
For the above Crab Nebula parameters, eq. (10) 
gives $\nu_c=4.2\times 10^{12}$ Hz
for $\epsilon_B=0.5$ and eq. (11) gives $\nu_m=1.3\times 10^{14}$ Hz.
The fact that $\nu_c<\nu_m$ implies that a spectrum of the form $L_{\nu}
\propto \nu^{-0.5}$ should occur between these frequencies.
Fig. (13) of KC84b shows that such a region does appear to be present
at optical wavelengths.

Table 1 gives a summary of well-established X-ray nebulae around
pulsars.
Kawai et al. (1998) list 6 additional nebulae, but 
Becker et al. (1999) and 
Pivovaroff, Kaspi, \& Gotthelf (2000) do not confirm the
diffuse emission seen with {\it ASCA} in  5 of the cases.
The value of $\dot E$ is determined from $\dot E=I\Omega\dot\Omega$,
where $I=10^{45}$ g cm$^2$ is the assumed neutron star moment of inertia
and $\Omega$ is the pulsar spin rate.
The X-ray luminosities are mostly from the {\it Einstein} observatory
and cover the energies 0.2--4 keV.
Several of the photon indices are from {\it ASCA} because of its
relatively broad energy coverage.
Two spectral indices are listed for G11.2--0.3; this remnant shows
thermal and nonthermal emission and the first one fits the lines with
Gaussians, while the second uses a non-equilibrium ionization model
(Vasisht et al. 1996).

N158A has properties that are very close to the Crab Nebula, but
N157B has a lower value of $L_x/\dot E$.
The observed photon index for N157B implies $p\approx 3$ if it is in
the cooling regime.
With $p=3$, $\epsilon_e=\epsilon_B=0.5$,
 and $\gamma_w$ and $r_s$ as for the Crab Nebula,
eq. (\ref{lum}) yields $L_{x}({\rm 0.2-4~keV})=3\times 10^{36}\ergs$.
The lower efficiency is naturally produced; the discrepancy with the
observed luminosity is not a problem because the
higher value of $p$ leads to a greater sensitivity to the model
parameters.

The compact nebula around
the Vela pulsar has   an especially low value
of $L_x/\dot E$ and a relatively flat spectrum.
The value of $L_x$ I have listed here is lower that often quoted because
I have taken a distance of 250 pc, as advocated by Cha, Sembach, \& Danks (1999)
and others; even at 500 pc, it is a low luminosity object.
The properties of the nebula indicate that it is not in the cooling
regime, i.e. $\nu_c>\nu_x$ for Vela.
de Jager, Harding, \& Strickman
 (1996) note that the nebular luminosity out to 0.4 MeV
is 0.1\% $\dot E$, implying a high value of $\nu_c$ ($\ga 10^{20}$ Hz).
The value of $r_s$ for Vela can be estimated from the size of the
compact nebula, which appears to be a bow shock nebula (Markwardt \&
\"Ogelman 1998).
At 250 pc, the radius is $2\times 10^{17}$ cm.
The proper motion of the pulsar (\"Ogelman, Koch-Miramond, \& Auli\`ere
 1989) gives
a transverse velocity of $45\kms$ so that $t=r_s/v_p=4.4\times 10^{10}$ s.
The age of the Vela remnant is $\sim 10^4$ yr, so eq. (\ref{lum2}) should be
applicable and yields $L_{x}({\rm 0.2-4~keV})=2\times 10^{34}
(\epsilon_B/10^{-4})^{0.83}\ergs$.
The non-cooling behavior requires $\nu_c\ga 10^{20}$ Hz or
$\epsilon_B\la 10^{-4}$.
In a similar argument, de Jager et al. (1996) set $t=r\sqrt{3}/c$ as the
escape time and thus found a  limit on the magnetic 
field close to equipartition.
The model here assumes an MHD flow through a bow shock region, which gives
a longer residence time for the particles.
The present model implies considerable 
variation in $\epsilon_B$ among nebulae;
such variation has little effect on $L_{\nu}$ in the cooling case
(see eq. [\ref{lcrab}]).

CTB 80 appears to be in the same class as the Vela nebula; it has a bow
shock structure (Safi-Harb, \"Ogelman, \& Finley 1995) and has a relatively
hard spectrum and low efficiency of conversion to X-ray emission (Table 1).
W44 also has a bow shock nature (Harrus, Hughes, \& Helfand 1996),
but the spectrum and X-ray luminosity are too uncertain to draw clear
conclusions.

The other 2 objects in Table 1 have $L_x/\dot E \sim 0.01$.
If the nebulae have $p\approx 2.2$ and rapid cooling in the X-ray
regime, the weak sensitivity to the parameters in eq. (\ref{lcrab})
makes it difficult to reduce $L_x/\dot E$ by the required
 factor of 5 compared to
the Crab.
However, MSH 15-52 appears to have a flatter X-ray spectrum than the Crab.
du Plessis et al. (1995) find that there is steepening of the X-ray spectrum,
consistent with $p=2.2$ and a synchrotron cooling break at $\sim 6$ keV.
A cooling time $\ga$ the age of the remnant is also indicated by the large
size of the X-ray nebula (Seward et al. 1984b).
The lack of rapid cooling can account for the lower efficiency of X-ray
production.
G11.2--0.3 may also have a relatively flat spectrum (Vasisht et al. 1996),
but there is insufficient information to draw firm conclusions on this source.

The observed spectral index of the Crab Nebula is consistent with
injection at the shock front with $p=2.2$.
This value of $p$ is consistent with expectations for particle acceleration
in a highly relativistic shock front; Bednarz \& Ostrowski (1998) found
that $p\rightarrow 2.2$ in the limit of a high Lorentz factor shock wave.
However, N157B does have a steeper spectral index, which suggests
$p\approx 3$, and $p=2.2$ does not appear to apply to some of the other objects.
In the context of the present model, detailed spectral study of X-ray
pulsar nebulae will be valuable for the determination of the particle
spectrum created in a relativistic shock front.
The same uncertainty is present in models of GRB (gamma-ray burst) afterglows.
The model presentation here was chosen to parallel that for GRB afterglows
(e.g., Sari et al. 1998) in order to show the similarities.

The present models can be used to predict the spin-down power of
the putative pulsars in wind nebulae where a pulsar has not yet been observed.
The value of the models over a purely empirical $L_x - \dot E$ relation
is that additional information about the nebula should allow a more
precise determination.
An excellent example is provided by the {\it Chandra} observations
of G21.5--0.9 (Slane et al. 2000).
Slane et al. find  a photon index $\alpha_X\approx 1.5$
for an inner core and
 $\alpha_X\approx 2.0$ for the whole wind nebula.
The spectral steepening is that expected for synchrotron burn-off.
A model like that for the Crab Nebula should apply, with high efficiency of
conversion of spin-down power to X-ray emission.
  For a distance of 5 kpc, the observed
  X-ray luminosity is $2.1\times 10^{35}\ergs$
  (Slane et al. 2000), from which I estimate $\dot E\approx
5\times 10^{36}\ergs$, as compared to $\dot E\approx
3.5\times 10^{37}\ergs$ deduced by Slane et al. (2000).
Detection of a pulsar will allow a test of this prediction.

\acknowledgments
Support for this work was provided in part by NASA grant NAG5-8130.

\clearpage
\begin{table*}

\noindent{Table 1. Properties of Pulsars and their X-ray Nebulae }

\vspace{1cm}

\begin{tabular}{cccccccc}
\hline
PSR &  Supernova & Distance &  $\dot E$ &  Photon  & $L_x$  & $L_x/\dot E$ & {Refs.\tablenotemark{a}} \\
     &    Remnant  & (kpc)  & (ergs s$^{-1}$)& Index  & (ergs s$^{-1}$)
     &   & \\
\hline
J0537--69 &  N157B & 50  &  $4.8\times 10^{38}$    & 2.5  &  $5.6\times 10^{36}$  & 0.01 & (1),(2)  \\
B0531+21 &  Crab &  2 &  $4.7\times 10^{38}$    & $2.10\pm 0.01$  &  $2.3\times 10^{37}$  &0.05 & (3),(4) \\
B0540--69 &  N158A &  50 &  $1.5\times 10^{38}$    & 2.04$\pm 0.06$  &  $8\times 10^{36}$  &0.05 & (2),(4) \\
B1509--58 &  MSH 15-52 & 4.2 &  $1.8\times 10^{37}$    & $1.90\pm 0.04$  &  $1.5\times 10^{35}$  &0.01 & (5),(6) \\
B0833--45 &  Vela & 0.25 &   $7.1\times 10^{36}$    & $1.67\pm 0.01$  &  $1.7\times 10^{33}$  &0.0002 & (6),(7) \\
J1811--19 &  G11.2--0.3 & 5  &  $6.4\times 10^{36}$    & 1.4$^{+0.54}_{-0.68}$,1.80$^{+0.28}_{-0.54}$  &  $4.9\times 10^{34}$  &0.008 & (8),(9) \\
B1951+32 &  CTB 80 & 2.5  &  $3.8\times 10^{36}$    & $1.8\pm 0.1$  &  $6\times 10^{33}$  &0.002 & (6),(10) \\
B1853+01 &  W44 & 3 &   $4.3\times 10^{35}$    & $2.3^{+1.1}_{-0.9}$  &  $4^{+30}_{-3}\times 10^{33}$  &0.009 & (11) \\
 \hline
\end{tabular}
\tablenotetext{a}{The references to the data are: 
(1) Marshall et al. 1998; (2)  Wang \& Gotthelf 1998;
(3) Pravdo \& Serlemitsos 1981; (4) Seward, Harnden, \& Helfand 1984a;
(5)  Seward et al. 1984b; (6) Kawai et al. (1998);
(7) Harnden et al. (1985);
(8) Torii et al 1999; (9) Vasisht et al. 1996;
(10) Safi-Harb, \"Ogelman, \& Finley 1995;
(11) Harrus et al. 1996}

\end{table*}


\clearpage



\begin{thebibliography} {}



\bibitem[]{}
Begelman, M. C. 1998, \apj, 493, 291

\bibitem[]{}
Becker, W., Kawai, N., Brinkmann, W., \& Magnani, R. 1999, \aap, 352, 532

\bibitem[]{}
Bednarz, J., \& Ostrowski, M. 1998, Phys. Rev. Lett., 80, 3911

\bibitem[]{}
Cha, A. N., Sembach, K. R., \& Danks, A. C. 1999, ApJ, 515, L25

\bibitem[]{}
de Jager, O. C., Harding, A. K., \& Strickman, M. S. 1996, \apj, 460, 729

\bibitem[]{}
du Plessis, I., de Jager, O. C., Buchner, S., Nel, H. I., North, A. R.,
Raubenheimer, B. C., \& van der Walt, D. J. 1995, \apj, 453, 746

\bibitem[]{}
Harnden, F. R., Jr., Grant, P. D., Seward, F. D., \& Kahn, S. 1985, 
\apj, 299, 828

\bibitem[]{}
Harrus, I. M., Hughes, J. P., \& Helfand, D. J. 1996, \apj, 464, L161

\bibitem[]{}
Kawai, N., Tamura, K., \& Shibata, S. 1998, in Neutron Stars and Pulsars:
Thirty Years after the Discovery, eds. 
N. Shibazaki, N. Kawai, S. Shibata, \& T. Kifune (Tokyo: Universal Acad.),
449

\bibitem[]{}
Kennel, C. F., \& Coroniti, F. V. 1984a, ApJ, 283, 694 (KC84a)

\bibitem[]{}
Kennel, C. F., \& Coroniti, F. V. 1984b, ApJ, 283, 710 (KC84b)

\bibitem[]{}
Kundt, W., \& Krotscheck, E. 1980, \aap, 83, 1

\bibitem[]{}
Markwardt, C. B., \& \"Ogelman, H. B. 1998, Mem. Soc. Astr. Ital., 69, 927

\bibitem[]{}
Marshall, F. E., Gotthelf, E. V., Zhang, W., Middleditch, J., \& Wang, Q. D.
1998, \apj, 499, L 179

\bibitem[]{}
\"Ogelman, H. B., Koch-Miramond, L., \& Auri\`ere, M. 1989, \apj, 342, L83

\bibitem[]{}
Pivovaroff, M. J., Kaspi, V. M., \& Gotthelf, E. V. 2000, \apj, 528, 436

\bibitem[]{}
Pravdo, S. H., \& Serlemitsos, P. 1981, \apj, 246, 484

\bibitem[]{}
Rees, M. J., \& Gunn, J. E. 1974, MNRAS, 167, 1



\bibitem[]{}
Safi-Harb, S., \"Ogelman, H., \& Finley, J. P. 1995, \apj, 439, 722

\bibitem[]{}
Sari, R., Piran, T., \& Narayan, R. 1998,
\apj, 497, L17

\bibitem[]{}
Seward, F. D., \& Wang, Z. 1988, ApJ, 332, 199

\bibitem[]{}
Seward, F. D., Harnden, F. R., Jr., \& Helfand, D. J. 1984a, \apj,
287, L19

\bibitem[]{}
Seward, F. D., Harnden, F. R., Jr., Symkowiak, A., \& Swank, J. 1984b, \apj,
281, 650

\bibitem[]{}
Slane, P., Chen, Y., Schulz, N. S., Seward, F. D., Hughes, J. P., \&
Gaensler, B. M. 2000, \apj, 533, L29

\bibitem[]{}
Torii, K., Tsunemi, H., Dotani, T., Mitsuda, K., Kawai, N., Kinugasa, K.,
Saito, Y., \& Shibata, S. 1999, \apj, 523, L69

\bibitem[]{}
Vasisht, G., Aoki, T., Dotani, T., Kulkarni, S. R., \& Nagase, F.
1996, \apj, 456, L59

\bibitem[]{}
Wang, Q. D., \& Gotthelf, E. V. 1998, \apj, 494, 623




%


\end{thebibliography}
\end{document}